# Parallel Neuron Groups in the *Drosophila* Brain

Robert Worden

Active Inference Institute, Crescent city, CA, USA

rpworden@me.com

Draft 0.8; December 2025

Abstract:

The full connectome of an adult *Drosophila* enables a search for novel neural structures in the insect brain. I describe a new neural structure, called a Parallel Neuron Group (PNG).

Two neurons are called parallel if they share a significant number of input neurons, and a significant number of output neurons. I define a quantitative measure of the parallel match between two neurons, which has a range 0.0 – 1.0. Most pairs of neurons in the *Drosophila* brain have very small parallel match. However, there are about twenty groups of neurons for which any pair of neurons in the group has a high match. These are the parallel neuron groups.

Parallel groups are statistically distinguished from other neurons; they are uncommon (in total comprising only about 1000 out of 65,000 neurons), and have distinctive properties. Group sizes range from about 150 neurons down to about 10 neurons. There are groups in the right mushroom bodies, the antennal lobes, the lobula, and in two central neuropils (GNG and EB).

15 of the 20 identified groups do not have lateral symmetry. A group usually has one major input neuron, which inputs to all the neurons in the group, and a small number of major output neurons. These major input and output neurons are laterally asymmetric. Most groups consist of excitatory neurons, with weak cross-links between them. A small group in the ellipsoidal body is inhibitory and has strong cross-coupling between its neurons.

Parallel neuron groups present puzzles, such as: what does a group do, that could not be done by one larger neuron? Do all neurons in a group fire in synchrony, or do they perform different functions? Why are they laterally asymmetric? These may merit further investigation.

# 1. Introduction

The publication of the full connectome of an adult *Drosophila* [Lin et al, 2024] gives opportunities to look for novel neural structures in the insect brain. Here I describe a new structure called a **Parallel Neuron Group**.

Two neurons are called parallel if they share a large proportion of their input neurons, and a large proportion of their output neurons. Parallel neurons are interesting because, having similar inputs and outputs, they might be expected to be doing the same thing. They raise the questions: how are parallel neurons different from each other? Why do parallel groups exist?

For each neuron, one can measure how many parallel neurons it has. For the great majority of neurons in the *Drosophila* brain, this number is less than 5, and is usually zero. However, for a small minority (fewer than 1000 neurons, out of 65,000) there are up to 160 parallel neurons. When a neuron has many parallel neurons, most of the parallel neurons are parallel to each other. A **parallel group** is a set of neurons, with high average match between any neuron and any other neuron in the group. There are three groups with size greater than 80, and about 15 smaller groups. Fewer than 1000 out of 65,000 neurons are in a parallel group. These groups have distinctive properties, which may make them worth further study:

- Statistically, they stand out clearly from other neurons in the brain; there can be little doubt that they are a real feature of the *Drosophila* brain.
- In most of the groups (but not all) there are only weak cross-connections between neurons in the group; neurons in a group appear to act independently of each other.
- For all groups but one, all the neurons in a group are excitatory
- In nearly all groups, each neuron in the group is in a 2-neuron feedback loop with one other neuron, which I call the master neuron of the group. Several groups share the same master neuron. For most groups, the master neuron is inhibitory, making a negative feedback loop. The master neurons are amongst the largest (in terms of synapses) in the *Drosophila* brain.
- There is a question of why so many neurons all have the same input and output connections as each other, so might be expected to do the same thing. This appears to be an inefficient use of neurons. Are the firing patterns of neurons in a group similar or different?
- Nearly all groups have lateral asymmetry, spanning one set of neuropils, but with no mirror group on the other side of the brain. This adds to the mystery of what parallel groups do.

Lateral asymmetry has been reported in the *Drosophila* brain, mainly in a small neuropil (the Asymmetric Body, AB) in the central complex [Pascual et al 2004; Wolff & Rubin 2018, Zhang et al 2024, Abubaker et al 2024]. Parallel groups and their master neurons extend the known asymmetries to the mushroom bodies, the antennal lobes, and the optic lobes.

The properties of parallel groups are distinctive. I suggest that they are not just a curiosity, but may merit further investigation and shed new light on the puzzles of the brain. They are a challenge both for experimental and theoretical neuroscience, to measure their dynamic properties, and to give an account of what they are doing. It is also possible, given recent advances in mammalian connectome data [Lee, Dubuc et al. 2025], to ask whether there are parallel groups in the mammalian brain.

In section 2, I define the concept of parallel groups an describe the method of finding them. Section 3 is a table of parallel groups with their main input neurons. Sections 4-7 describe and illustrate specific groups. Section 8 concludes.

# 2. Methods and Definitions

The *Drosophila* connectome data of [Lin et al 2024] can be downloaded from the flywire codex database at https://codex.flywire.ai/?dataset=fafb . They are the source of the data analysed and used in this paper.

Two main data sources in the flywire codex database have been used:

1. A large (11GB) .feather data file giving the coordinates, neurotransmitters, neuron ids, and neurotransmitters of all 130 million synapses. I have used Python utilities to split the file into approximately 1900 comma-separated value (.csv) files of 64,000 lines each.
2. A folder of 65,000 files in the SWC text format, each one giving the coordinates of the sections of one neuron. The file names are full neuron identifiers of 18 digits.

I have written Java programs to perform a variety of statistical analyses on these files, to find the parallel groups (as defined in the mathematical definitions which follow in this section), and to make graphical displays of the neurons and neuropils. These programs are available on request.

The purpose of the analysis was an open-ended investigation of the *Drosophila* brain, to find any interesting or puzzling phenomena, with few preconceptions or null hypotheses. Two null hypotheses which became prominent during the investigation were:

- Every neuron in the brain serves some distinct purpose
- The brain has bilateral symmetry, reflecting the bilateral symmetry of the *Drosophila* body and of its experienced world

**Definition of Parallel Neuron Pairs**

Two neurons may have inputs from overlapping sets of neurons. If the number of synapses from neuron b to neuron a is $n_{ba}$, then for a neuron b, the set of all $n_{ab}$, for all other neurons a in the brain, is a vector in a space of dimension 64,000, having 64,000 independent components. These vectors are denoted by bold letters:

$\mathbf{b} = \{n_{ab}\}$ for all neurons a

The scalar product of these vectors is:

$\mathbf{b}.\mathbf{c} = \sum_a n_{ab} n_{ac}$

These scalar products can be zero (for two neurons with no feeder neurons in common), or can be large integers. For instance

$\mathbf{b}.\mathbf{b} = \sum_a n_{ab}^2$

This number can typically be as large as a thousand, if a neuron has inputs from 100 other neurons, some of which have many synapses to it (e.g. $n_{ab} = 50$, or 500 after correction).

The **input match** between two neurons b and c is defined as

$m_{bc} = \mathbf{b}.\mathbf{c} / \sqrt{(\mathbf{b}.\mathbf{b} * \mathbf{c}.\mathbf{c})}$

$m_{bc}$ is in the range 0.0 – 1.0. $m_{bc}$ is a measure of the input similarity of two neurons. When $m_{bc}$ is large, the two vectors are approximately parallel in the large-dimensional space.

There is an analogous measure of the **output match** $n_{bc}$ of any two neurons, and the **overall match** is defined by

$p_{bc} = m_{bc} * n_{bc}$

$p_{bc}$ is in the range 0.0 – 1.0, and is a measure of the overall match (input and output) of two neurons, and measures the extent to which their inputs are parallel and their outputs are parallel.

For two neurons b and c randomly selected in the fruitfly brain, $p_{bc}$ is expected to be zero or very small. However, for one neuron b, if you look for the neuron c with the best match to b, there is an appreciable number of large matches. A histogram of these best matches, for a sample of 1000 neurons, is shown in figure 1:

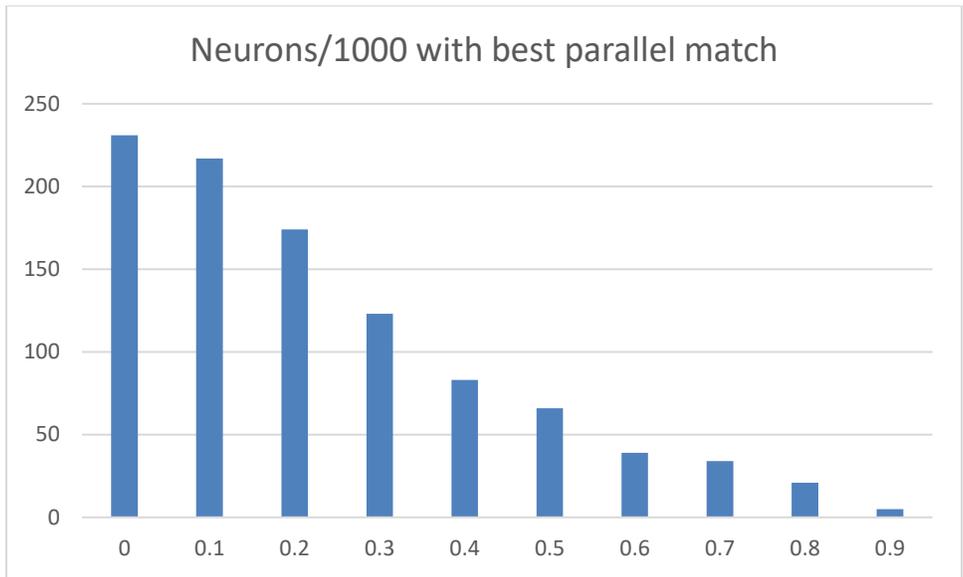

*Figure 1: Best parallel match of neurons with any other neuron in the brain, for a sample of 1000 neurons*

Figure 1 implies that the number of neuron pairs with $p_{bc} \geq 0.6$ is 99/1000, approximately 10% of the total. If we define $p_{bc} \geq 0.6$ as the criterion for a matching pair of neurons, then approximately one neuron in ten is part of a parallel neuron pair. Parallel neuron pairs on their own are not the focus of this paper.

**Definition of Parallel Neuron Groups**

Many neurons have one parallel neuron, but few of these have two parallel neurons, even fewer have three, and so on. We can compute a probability distribution of how many parallel neurons a randomly chosen neuron has, and the result is shown in figure 2.

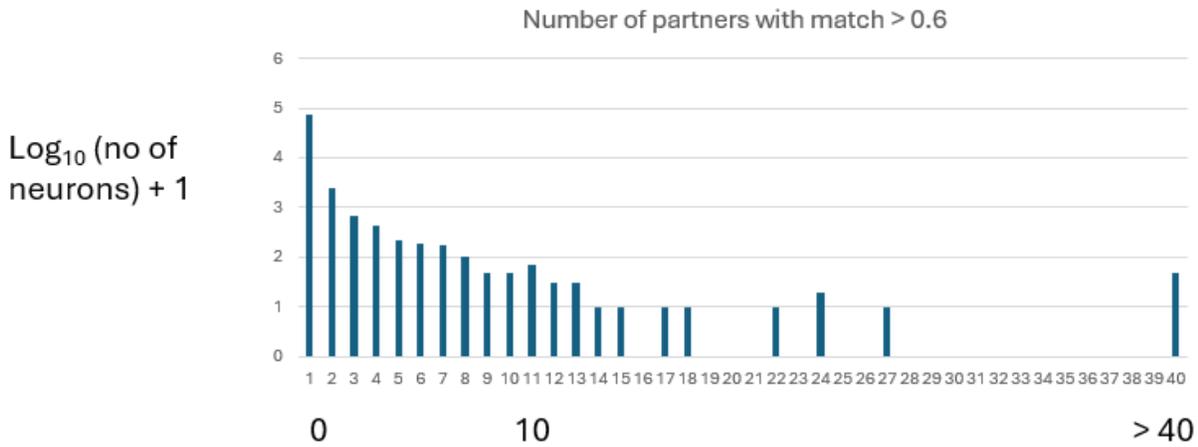

| 0 | 1 | 2 | 3 | 4 | 5 | 6 | 7 | 8 | 9 | 10 | 11 | 12 | 13 | 14 | 15 | 16 | 17 | 18 | 19 | 20 | 21 | 22 | 23 | 24 | 25 | 26 | 27 | 28 | 29 | 30 | 31 | 32 | 33 | 34 | 35 | 36 | 37 | 38 | >38 |
|---|---|---|---|---|---|---|---|---|---|---|---|---|---|---|---|---|---|---|---|---|---|---|---|---|---|---|---|---|---|---|---|---|---|---|---|---|---|---|---|
| 7491 | 251 | 70 | 43 | 22 | 19 | 18 | 10 | 5 | 5 | 7 | 3 | 3 | 1 | 1 | 0 | 1 | 1 | 0 | 0 | 0 | 1 | 0 | 2 | 0 | 0 | 1 | 0 | 0 | 0 | 0 | 0 | 0 | 0 | 0 | 0 | 0 | 0 | 0 | 5 |

*Figure 2: Distribution of the number of parallel neurons of any neuron, plotted from a sample of 8000 randomly selected neurons. The graph is logarithmic, and the table shows the numbers of neurons with N parallel neurons, as a function of N*

This histogram has been plotted with a random sample of 8000 neurons. Of these, 7491 neurons have no parallel neuron. Although the initial distribution of the number of parallel neurons tails off approximately exponentially (251,70,43,22…), the distribution has a long tail, including five neurons each of which has more than 38 parallel neurons. This long tail is a statistically significant indication that something else is going on – that a small minority of neurons are members of some **parallel group**, in which any neuron in the group has a high parallel match with any other neuron in the group.

I define a parallel group as follows:

A **parallel group** is a set of neurons $\{n_a, n_b,…\}$ for which the average of the matches $p_{ab}$ of any neuron a with all the other neurons b in the group is greater than 0.6.

It is then possible to search programmatically for all parallel groups greater than a minimum size, which I have set to 9. With this criterion, I have found about 30 groups in the size range 9 – 170 neurons, totalling 1000 neurons, which are described in the rest of this paper.

Readers may be concerned that the criterion for a parallel group is arbitrary. Why the particular definition of the match $p_{ab}$ between two neurons, and why the threshold 0.6? Like most science, this is work in progress. It is like a chemist, who has found out how to synthesize some new compound, but not yet to purify it to a crystalline form. For instance, as the threshold is varied between 0.5 and 0.7, the sizes of most groups vary. In future a better definition of a parallel group may be found, leading to stable group sizes; or possibly, there is no neat crystalline form, and parallel groups just have fuzzy boundaries. Meanwhile, the groups have enough interesting properties to make them worth describing.

A prominent feature of most parallel groups is that all neurons in the group have strong inputs from a small number of large neurons, and output to a few large output neurons. These input and output neurons are the largest neurons in the *Drosophila* brain (in terms of total number of detected synapses), as will be described in the next section. Sometimes the contribution of one input neuron is so large as to dominate the contributions of all other neurons, which raises a concern. Is the parallel group totally defined by its relation to the dominant input neuron? If that neuron was removed from the definition of the scalar products in the group, would the neurons in the group still be parallel to one another? If not, the groups might be defined not so much by similarity of neuron in the group, but by the dominance of one input neuron.

To explore this concern, I have defined the **master input neuron** of a group to be the neuron whose total number of input synapses to the group is the largest; and the **reduced match** between two neurons in a group, to be their overall match with the master input neuron removed from the sums in the scalar products. I shall discuss the mean overall match and the mean reduced match for each group.

Parallel groups are found in only a few neuropils. The abbreviations of the neuropil names are those used in [Lin et al 2024]. The full identifiers of the neurons in the parallel groups, and of the input and output neurons of parallel groups, are provided in supplementary material.

## 3.  Major Input/Output Neurons and Lateral Asymmetry

For most parallel groups there are a few large neurons (measured in terms of their total number of synapses) which dominate the inputs to and outputs to neurons in the parallel group. These major input and output neurons are the largest neurons in the *Drosophila* brain. The 30 largest neurons are shown, in descending order of their size, in table 1:

| rank | id | excit / inhibit | main input neuropil | main output neuropil | parallel groups | total synapses detected | input neurons | output neurons |
| --- | --- | --- | --- | --- | --- | --- | --- | --- |
| 1 | 720575940628908548 | m | LO_L | LO_L | lo1 lo2 | 211691 | 4736 | 4549 |
| 2 | 720575940613583001 | - | MB_ML_R | MB_CA_R | mb1 - mb10 | 160407 | 1466 | 1439 |
| 3 | 720575940625952755 | - | LO_L | LO_L | lo3 lo5 | 127352 | 2547 | 3486 |
| 4 | 720575940628307026 | - | LOP_R | LOP_R | | 90794 | 2103 | 3997 |
| 5 | 720575940619991862 | - | LOP_L | LOP_L | lo1 lo3 | 88641 | 1549 | 1577 |
| 6 | 720575940628038808 | + | ME_R | ME_R | | 72699 | 698 | 4043 |
| 7 | 720575940650935673 | - | LO_R | LO_R | | 70529 | 1693 | 2044 |
| 8 | 720575940628069501 | - | LO_L | LO_L | | 66344 | 1649 | 2027 |
| 9 | 720575940618308825 | + | AL_L | MB_CA_L | al4 | 63601 | 169 | 616 |
| 10 | 720575940608545219 | - | AVLP_R | AVLP_R | lo4 | 61228 | 825 | 792 |
| 11 | 720575940618366843 | - | AVLP_L | AVLP_L | | 60486 | 356 | 383 |
| 12 | 720575940622523508 | - | LO_L | LO_L | | 60408 | 2468 | 1181 |
| 13 | 720575940626462014 | - | SPS_R | IPS_L | | 58565 | 150 | 1046 |
| 14 | 720575940627190556 | - | LOP_R | LOP_R | | 57001 | 1461 | 1094 |
| 15 | 720575940610584184 | - | AVLP_L | AVLP_L | | 55683 | 886 | 664 |
| 16 | 720575940612718563 | - | SPS_R | IPS_R | | 55233 | 114 | 1074 |
| 17 | 720575940636543668 | + | ME_R | ME_L | | 55032 | 2684 | 2112 |
| 18 | 720575940626403917 | + | ME_L | ME_R | | 54563 | 2537 | 2192 |
| 19 | 720575940630770042 | + | AL_L | MB_CA_L | al1 | 52242 | 170 | 482 |
| 20 | 720575940623888397 | m | ME_R | ME_R | | 51408 | 2172 | 1440 |
| 21 | 720575940624402173 | + | SMP_L | SMP_R | | 48350 | 650 | 801 |
| 22 | 720575940614710509 | + | AVLP_R | ME_R | | 48143 | 1290 | 2382 |
| 23 | 720575940640834211 | - | PVLP_L | PVLP_L | lo3 lo5 | 47954 | 564 | 475 |
| 24 | 720575940638900713 | - | PVLP_R | PVLP_R | lo4 | 47669 | 565 | 534 |
| 25 | 720575940632777320 | + | SPS_R | ME_R | lo2 | 46703 | 403 | 2924 |
| 26 | 720575940630810306 | - | LOP_L | LOP_L | lo1 | 46639 | 1283 | 983 |
| 27 | 720575940612976497 | + | GNG | ME_R | | 45350 | 567 | 2187 |
| 28 | 720575940621116807 | - | LO_R | LO_R | | 44640 | 2105 | 1582 |
| 29 | 720575940615743650 | - | LOP_L | LOP_L | lo1 lo3 | 44553 | 1318 | 1015 |
| 30 | 720575940625525740 | + | IPS_L | ME_L | lo2 | 44180 | 335 | 2676 |

*Table 1: The 30 largest neurons in the Drosophila brain, and their links to parallel neuron groups*

In the short neuropil names such as 'PVLP_R', the ending '_R' refers to the right side of the brain and '_L' refers to the left side of the brain. Lateral asymmetry in the table occurs when a large neuron in a '_R=>_R' neuropil is not balanced by a large neuron in a '_L=>_L' neuropil (a contralateral mirror equivalent). For instance, the neuron at row 6 (from ME_R to ME_R) is not balanced by any neuron ME_L to ME_L in the table.

It can be seen in the table that:

- The largest neurons have strong lateral asymmetry: For instance, the two largest neurons (with abbreviated ids 8548 and 3001) have no contralateral mirror equivalents in the table
- The largest neurons are strongly associated with parallel groups (13 out of the top 30 neurons have associated parallel groups)
- The associated parallel groups are in the lobula (lo1 – lo5), the right mushroom bodies (mb1- mb10) and the antennal lobes (al1 and al4). These parallel groups have strong lateral asymmetry.
- The largest neurons are mainly of two kinds:
    a. **balanced neurons** (where the number of neurons inputting to the large neuron is approximately the same as the number of with output from the large neuron)
    b. **transmitter neurons**, which connect to many more output neurons than input neurons
- Parallel groups are mainly associated with balanced neurons (and form tight feedback loops with them)

Between them, the largest neurons and the parallel groups show strong lateral asymmetry, outside the central complex of the brain. Previous evidence of lateral asymmetry [Pascual et al 2004; Wolff & Rubin 2018]. has been confined to two small neuropils in the central complex (called the asymmetric bodies, AB).

The large neurons in table 1 are not members of the rich club of *Drosophila* neurons described in [Lin et al 2024]. This is because the rich club is defined as those neurons in a range of neuron degree (where degree = number of input neurons + number of output neurons) which have more connections to each other than would be expected, as defined by a relative rich club coefficient greater than 1.0. The relative coefficient is defined with reference to a configuration graph model (CFG), a randomised model that preserves the distribution of degrees. Lin et al found a relative coefficient greater than 1 in the range of degree from 37 to about 80, whereas the neurons in table 1 are larger than this, with degree more than 600.

# 4. Parallel Groups in the Right Mushroom Bodies

There are ten parallel groups in the right-hand mushroom bodies and neighbouring neuropils. These groups are called mp1…mp10 (in descending order of their size) and they have sizes 165, 108, 41, 31, 27, 27, 26, 16, 12, and 10 neurons respectively. The largest group mb1 is shown in figure 3.

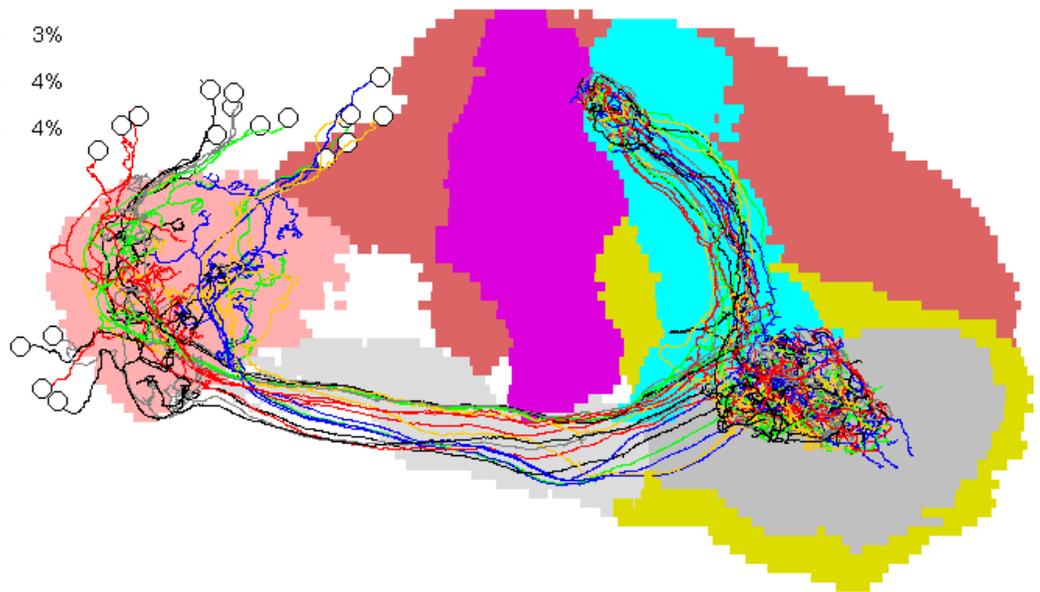

*Figure 3: The largest parallel group in the mushroom bodies*

In the figure:

1. The group is called mp1. It has 165 neurons, and a mean parallel match between pairs of neurons in the group of 69%, and a reduced match (when neuron 3001 is removed from the inputs and outputs of group neurons) of only 8%.
2. 10% of its input synapses are from within the group itself, and 0% of the neurons in the group are inhibitory.
3. 20 sample parallel neurons in the group are drawn – showing that all parallel neurons in the group have similar morphology. Their cell somas are shown as white circles.
4. Neuropil regions are shown coloured as in the key, which shows the percentages of synapses to parallel neurons (input and output) to parallel neurons in each neuropil
5. Neurons in the group start in the calyx, pass through the peduncle to the medial lobe (where they make most of their synapses, both input and output), and then on to the vertical lobe
6. There is a small minority of synapses in neighbouring neuropils – the crepine, the SIP and the SMP.
7. The leading input and output neurons to the group are shown in the order of the number of synapses they make to the group, each one showing the percentage of the neurons in the group that it connects to. The neuron percentages are not in exactly decreasing order, but in approximately descending order.
8. The leading neuron, here called 3001, has a full neuron id in the published connectome of 720575940613583001. It is an inhibitory neuron and is an input to all neurons in the group, and an output from all neurons in the group. Since they are excitatory neurons, this is a negative feedback loop.

It is possible to display the input and output neurons. The leading input neuron 3001 is the second largest neuron in the *Drosophila* brain, in terms of its total number of detected synapses. It has so many synapses that when neuron 3001 is displayed, it covers many of the neuropils in the group, as in figure 4:

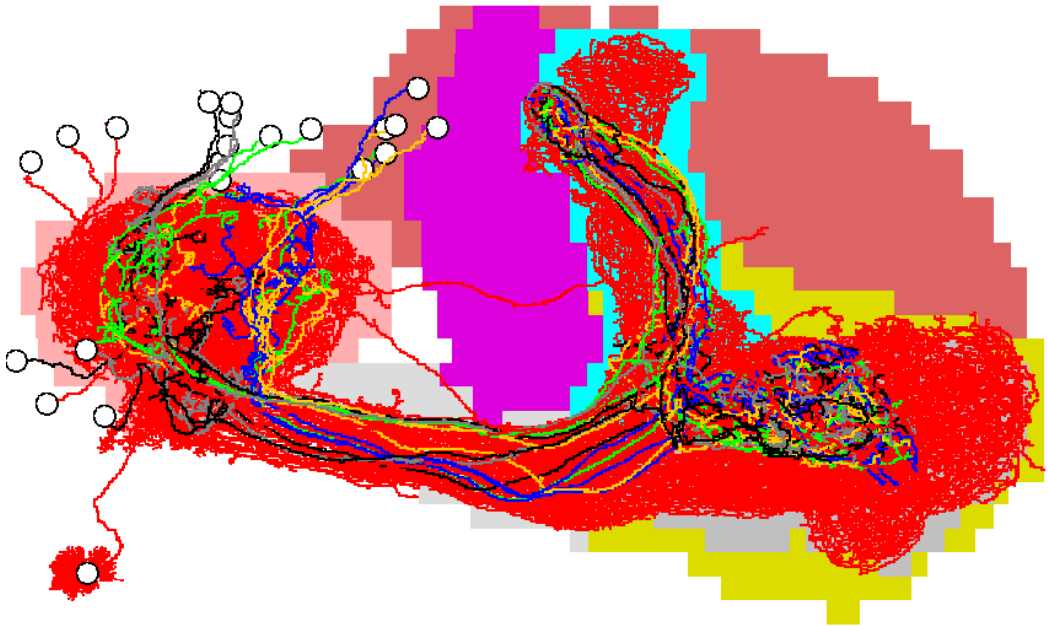

*Figure 4: the main input and output neuron 3001 of the group mb1, shown in red. The first 20 neurons in the group are superimposed in different colours.*

Neuron 3001 is strongly clustered in the neuropils (calyx, peduncle, median body and vertical body) occupied by the neurons in group mb1.

The large group mb1 has such a small reduced match (8%) when the input neuron 3001 is removed from the match, that one may question the nature of the group. Is it mainly defined by the links to 3001, rather than any other similarities between parallel neurons? I believe the group has more identity than that, for the following reasons:

- As can be seen from the figures, there are strong morphological similarities between neurons in the group
- There are other groups (mb2, mb5, mb6 and mb7), which also have small reduced matches (they are also dominated by 3001), but which are morphologically distinct from group mb1. So these groups are not solely defined by sharing the input neuron 3001.

The groups in the mushroom bodies with large reduced match (whose high matches are not dominated by one input neuron) are groups mb3, mb4, mb6, and mb10. There are morphological differences between some of them, and other distinctions, as shown in figure 5:

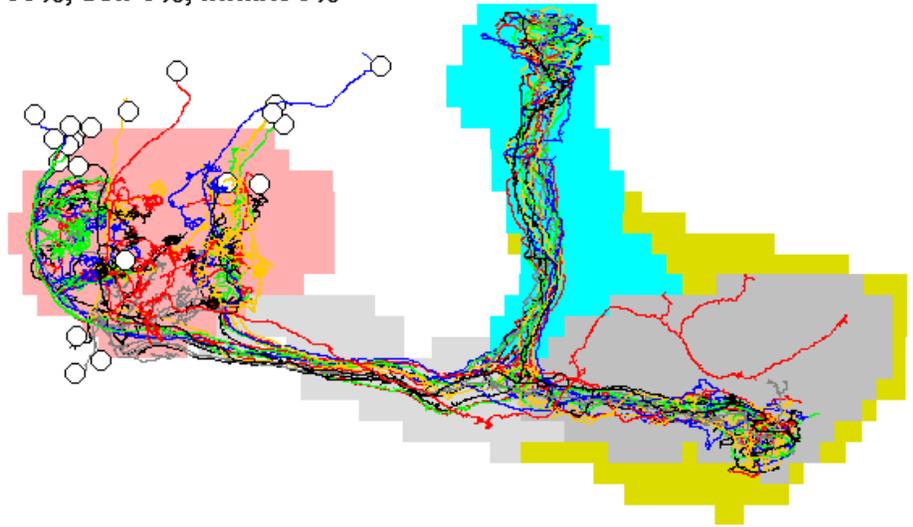

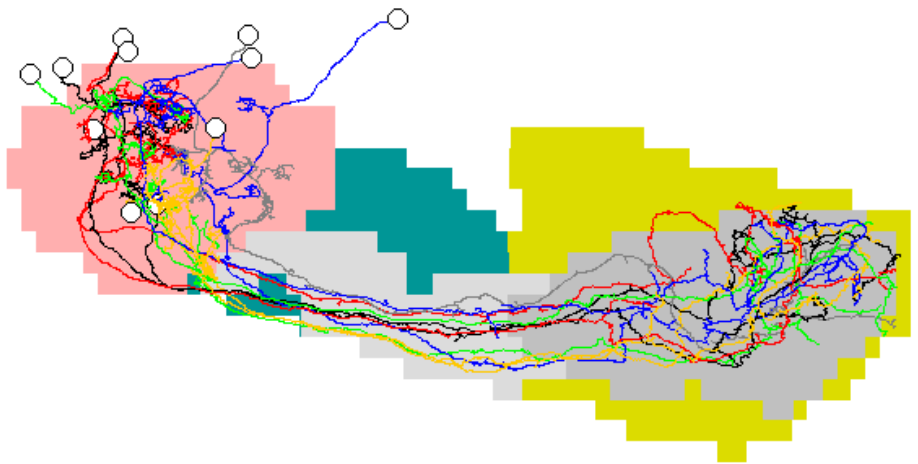

*Figure 5: Some differences between parallel groups in the mushroom bodies.*

While there are differences between some of the groups in the mushroom bodies, there should probably not be as many as 10 distinct groups; some of them could be merged. This is being investigated..

All the groups in the mushroom bodies have strong input and output links to the large inhibitory neuron 3001, so this negative feedback loop involves all the groups synchronously. It could be that 3001 acts as a kind of joint 'off' switch for all these groups, switching them all off at the same time. There is no contralateral equivalent to neuron 3001.

Parallel neurons in the mushroom bodies are classified as Kenyon cells in the flywire codex database [Lin et al 2024]; so they are associated with learning and memory.

These groups have strong lateral asymmetry. They are all in the right mushroom bodies, and there are no equivalent groups in the left mushroom bodies. [Abubaker et al 2024] have linked the Asymmetric Body (AB) in the central complex to olfactory learning in Drosophila. It would be interesting to see if the link from the AB to the mushroom bodies involves the parallel neuron groups in any way.

# 5. Parallel Groups in the Lobula

There are five parallel groups in the lobula, with names lo1 – lo5 and sizes 110, 92, 26, 13 and 13 neurons. Groups lo1, lo2, lo3 and lo5 are in the left lobula and are dominated by the largest neuron in the *Drosophila* brain (the neuron with short id 8548); if this neuron is removed, these four groups have low reduced match.

Group lo4 is the only group in the right lobula, and is shown in figure 5:

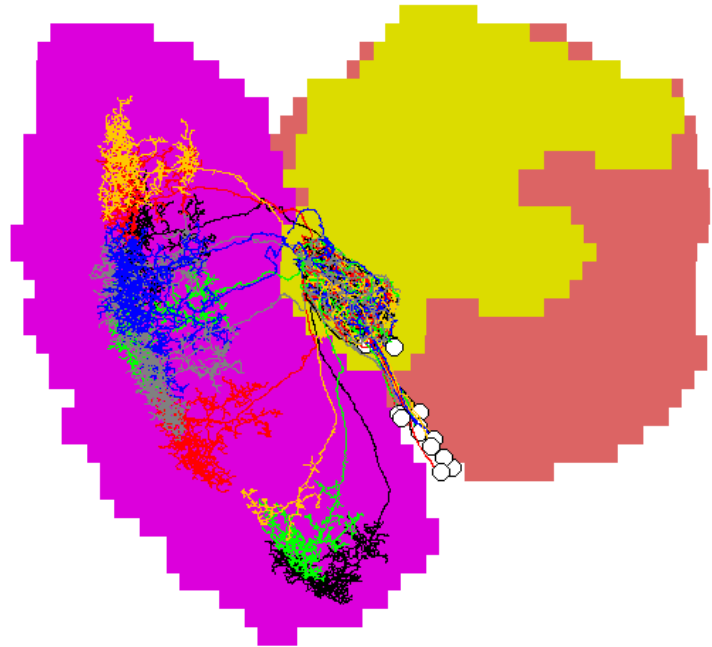

The other four groups, in the left lobula, are shown in figure 6.

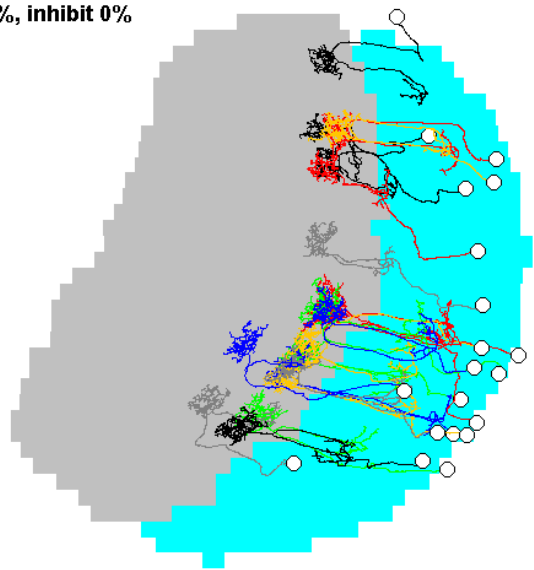

**lo2: size 92, match 68%, reduced 0%, self 0%, inhibit 0%**

| | npil | In | Out |
|---|---|---|---|
| ▓ (gray) | LO_L | 58% | 89% |
| ▓ (pink) | ME_L | 41% | 10% |

| Inputs | | Outputs | |
|---|---|---|---|
| 8548m | 100% | 8548m | 100% |
| 5740+ | 13% | 3818+ | 2% |
| 7418+ | 0% | 7418+ | 3% |
| 2404+ | 0% | 2404+ | 3% |
| 7950+ | 0% | 7950+ | 2% |
| 7357+ | 0% | 7357+ | 4% |
| 6874+ | 0% | 6874+ | 3% |
| 2089+ | 0% | 2089+ | 1% |

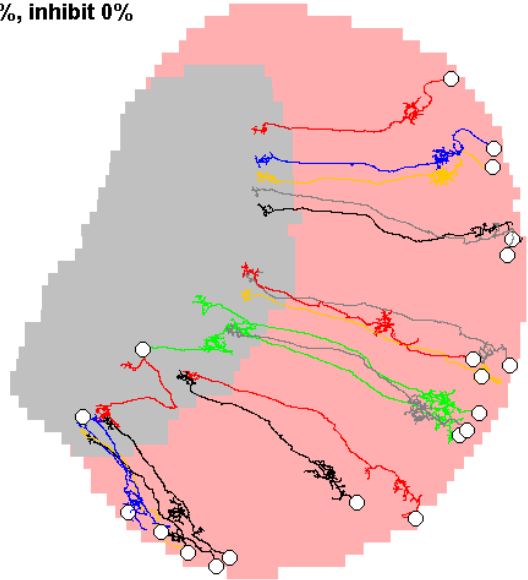

**lo3: size 26, match 66%, reduced 5%, self 9%, inhibit 0%**

| | npil | In | Out |
|---|---|---|---|
| ▓ (light gray) | PVLP_L | 26% | 85% |
| ▓ (gray) | LO_L | 43% | 9% |
| ▓ (cyan) | LOP_L | 30% | 4% |

| Inputs | | Outputs | |
|---|---|---|---|
| 4211- | 100% | 4211- | 100% |
| 2755- | 92% | 9153+ | 100% |
| 1862- | 69% | 1127+ | 100% |
| 5250+ | 53% | 9723- | 100% |
| 6467+ | 11% | 9783- | 100% |
| 3650- | 53% | 6980+ | 100% |
| 1185+ | 53% | 2487- | 100% |
| 0822+ | 50% | 4987- | 100% |

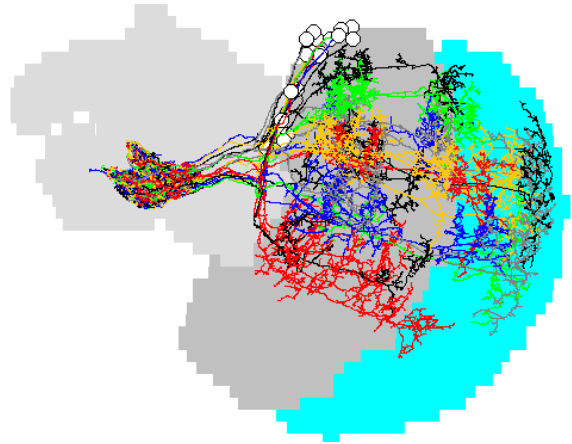

**lo5: size 13, match 69%, reduced 20%, self 2%, inhibit 0%**

| | npil | In | Out |
|---|---|---|---|
| ▓ (light gray) | PVLP_L | 28% | 93% |
| ▓ (gray) | LO_L | 71% | 6% |

| Inputs | | Outputs | |
|---|---|---|---|
| 4211- | 100% | 4211- | 100% |
| 2755- | 100% | 8500+ | 100% |
| 9457- | 100% | 5730+ | 100% |
| 7090- | 38% | 2483+ | 100% |
| 6716- | 53% | 6506+ | 100% |
| 4261+ | 30% | 8869+ | 100% |
| 0336- | 23% | 0414- | 92% |
| 3352+ | 38% | 0404- | 100% |

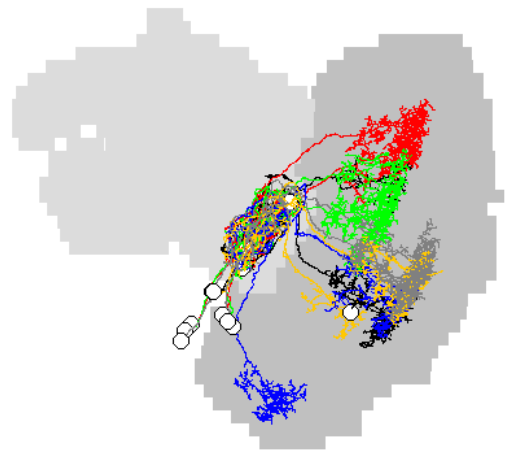

*Figure 6: Parallel groups in the left lobula. Different scales are used in different groups.*

As can be seen, these for groups are morphologically distinct from one another. Group lo5 is probably a mirror group to lo4, showing bilateral symmetry.

The two groups lo1 and lo2 have strong input and output connections to neuron 8548, which is the largest neuron in the *Drosophila* brain. It is excitatory, so these two groups are coupled in a positive feedback loop with that neuron. That large neuron has an interesting morphology, shown in close detail in figure 7:

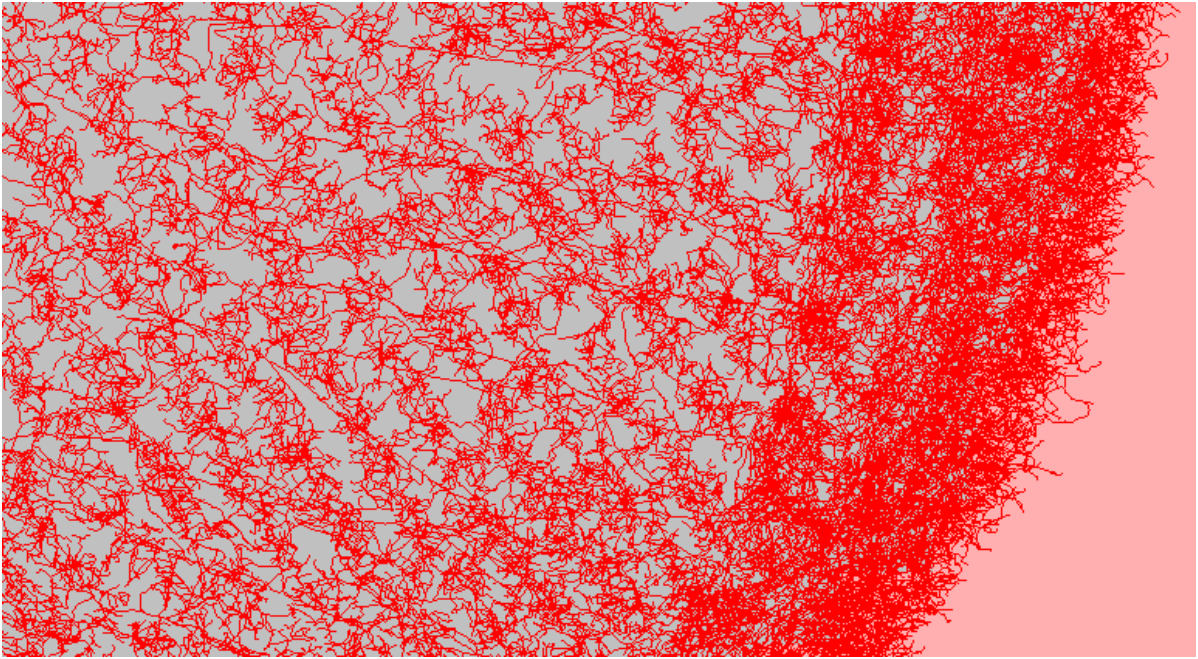

*Figure 7: detailed morphology of the largest neuron in the Drosophila brain (with short id 8548), in the left lobula*

This structure shows clear bands from top left to bottom right, with less clear orthogonal bands going from bottom left to top right, making a regular lattice. Why there should be this distinctive structure only in the left optic lobe appears to be a mystery.

A different view of the same neuron 8548 is shown in figure 8.

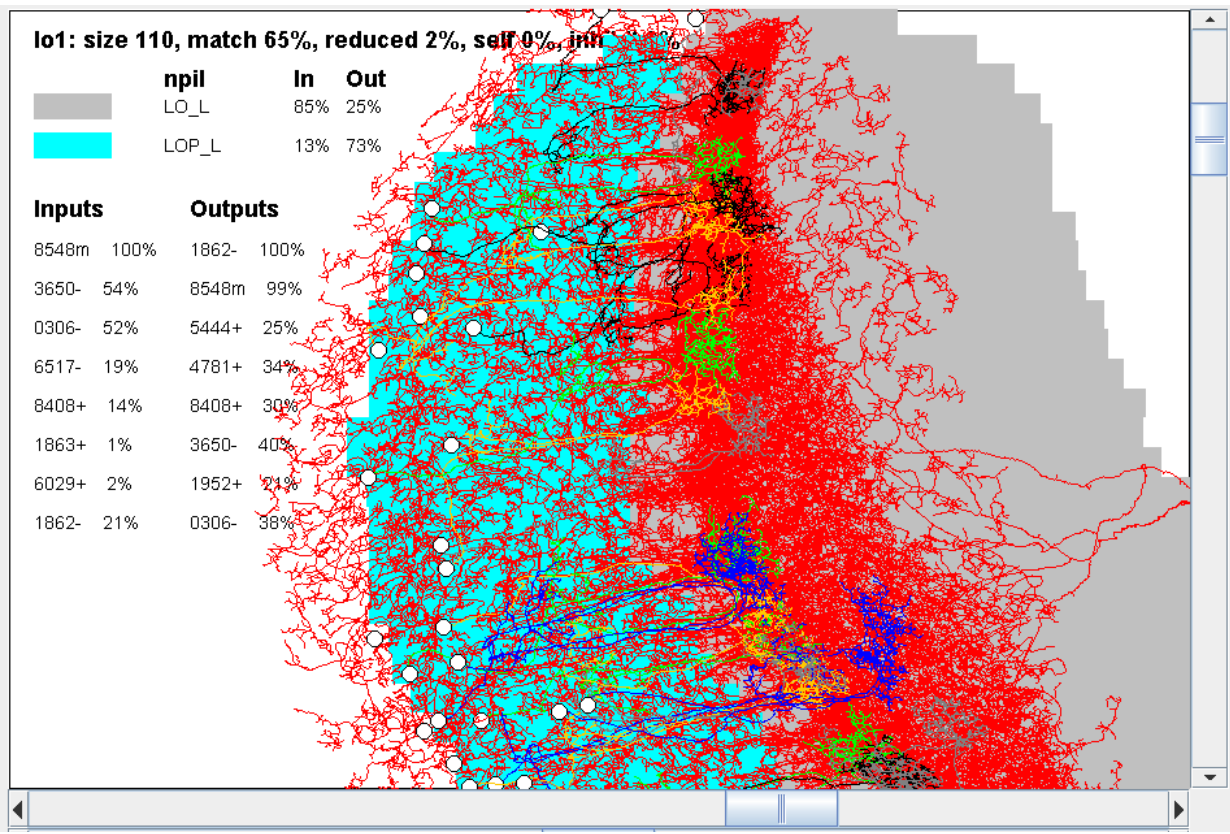

*Figure 8: another view of neuron 8548 (red), also showing neurons in group lo1 (different colours; somas shown as white circles)*

Figure 8 shows that neuron 8548 has distinctive bundles passing from the Lobula to the Lobula Plate, and that the neurons in group lo1 follow the paths of these bundles.

These neurons can also be examined at the flywire database of [Lin et al 2024]. For instance, the same neuron as in figures 7 and 8 is shown in figure 9:

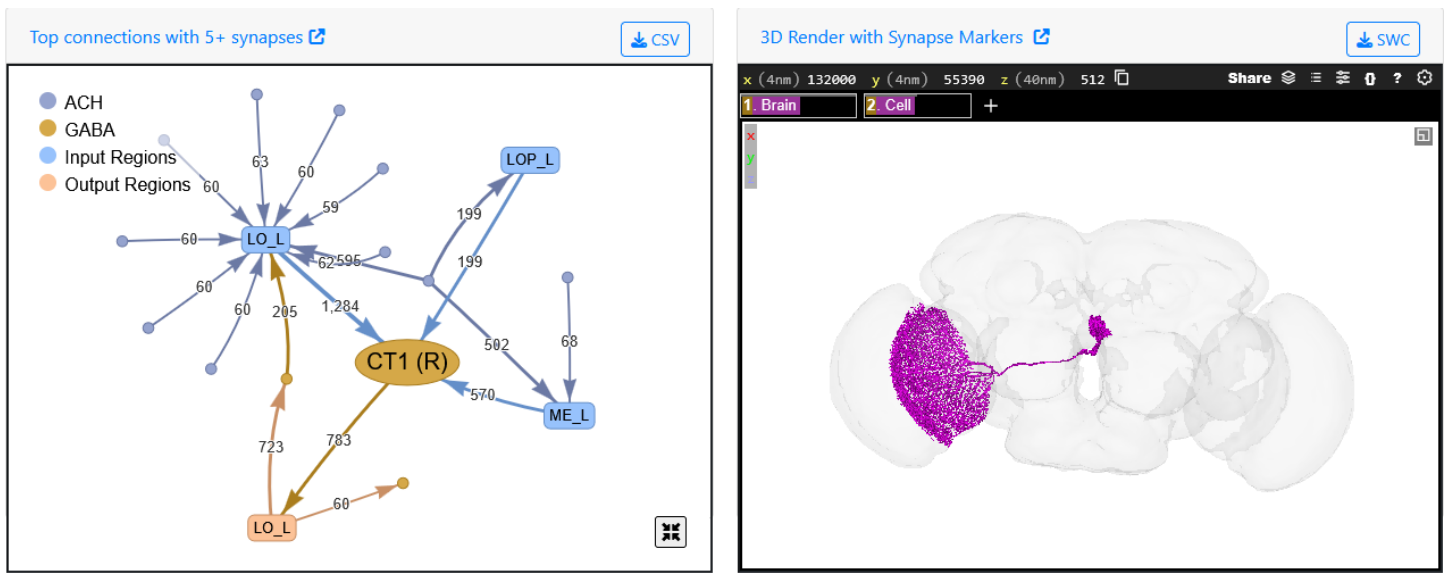

*Figure 9: the large neuron with short id 8548, as seen in the flywire codex database*

The strong left-right asymmetry of the parallel groups lo1- lo3 and their input/output neurons is puzzling, because visual stimuli on both sides of the animal are expected to be of the same general nature and importance; why is there neural circuitry in one optic lobe, but not in the other? It might be possible to look for some left-right asymmetry in *Drosophila* behaviour, where it prefers to use one side of its visual system in certain tasks.

## 6. Groups in the Antennal Lobes

There are 12 small groups in the antennal lobes, with sizes between 34 and 10. These are all confined to the antennal lobes and no other neuropils. They have a mixture of left-right lateral symmetry and asymmetry. The largest group is shown in figure 10.

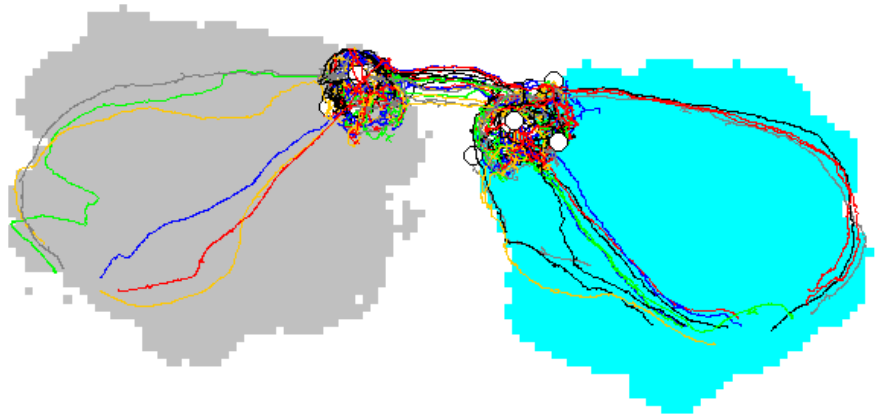

*Figure 10: the largest group in the antennal lobes*

This group has approximate left-right symmetry, and predominantly passes information from the right lobe to the left. The different neuron morphologies in the right lobe suggest that it may not be a pure group, but may be two groups.

Some groups which span both lobes are shown in figure 11:

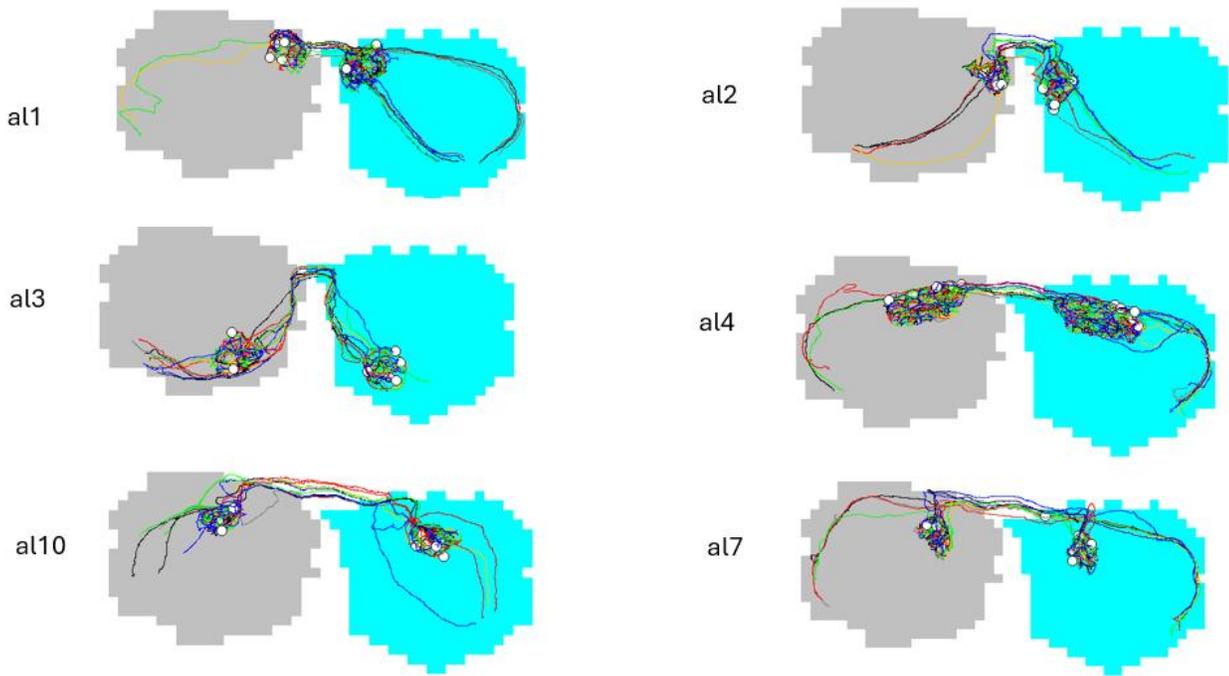

*Figure 11: Groups in the antennal lobes, some of which appear symmetric*

Although there are some possible merges or rearrangements of groups in figure 9, their morphological diversity is evident. Some more groups are shown in figure 12:

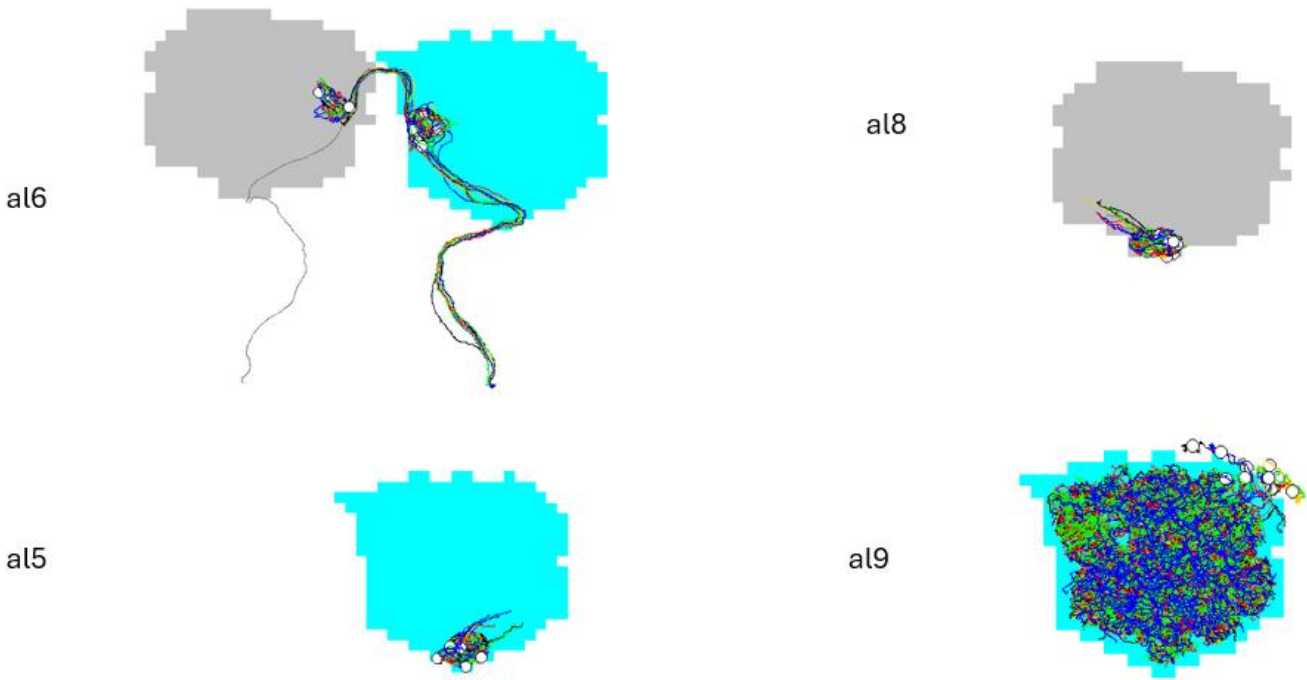

*Figure 11: other groups in the antennal lobes*

While some of the groups in the antennal lobes may be laterally symmetric, and groups al5 and al8 may be left-right mirror groups, group al9 stands on its own in one lobe, and has no mirror group.

## 7. Central Complex Parallel Groups

There are two small parallel groups in the central complex, in the gnathal ganglia and in the ellipsoidal body. The group in the gnathal ganglia is shown in figure 13:

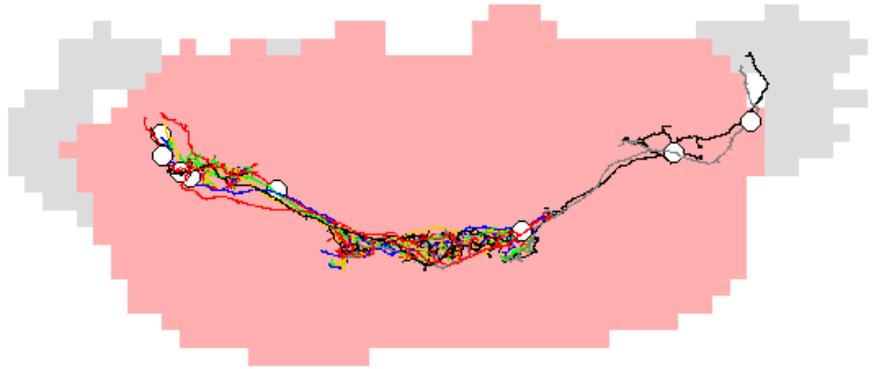

*Figure 13: parallel group in the gnathal ganglia.*

This group has a small extension in the saddle.

The group in the ellipsoidal body is shown in figure 14:

**eb1: size 14, match 75%, reduced 72%, self 68%, inhibit 100%**

| | npil | In | Out |
|---|---|---|---|
| 🟨 | EB | 93% | 96% |

| Inputs | | Outputs | |
|---|---|---|---|
| 2153- | 100% | 2153- | 100% |
| 4488- | 100% | 4488- | 100% |
| 5463m | 100% | 5280+ | 100% |
| 1236m | 100% | 2099+ | 100% |
| 1396+ | 7% | 5036+ | 100% |
| 3837m | 100% | 3837m | 100% |
| 4604- | 100% | 1236m | 100% |
| 2206m | 85% | 5463m | 100% |

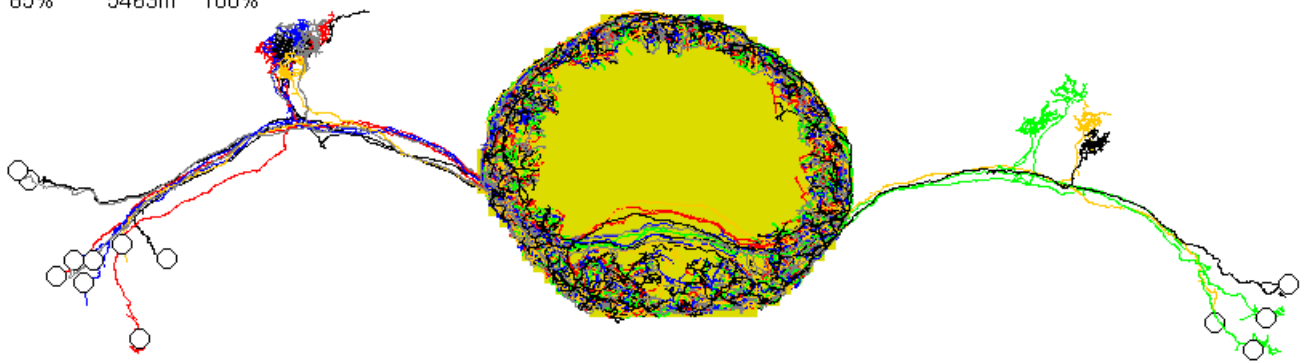

*Figure 14: parallel group in the ellipsoidal body*

This group follows the characteristic round shape of the *Drosophila* ellipsoidal body. It differs from all other parallel groups in two important respects:

- It consists entirely of inhibitory neurons (inhibit = 100%)
- The neurons in the group do not act in an independent manner; a large proportion of the inputs to any neuron in the group comes from other neurons in the same group (shown as 'self 68%').

## 8. Conclusion: Puzzles of Parallel Groups

Parallel neuron groups are a novel structure in the insect brain. They present some puzzles which may be topics for further research:

1. Why are so few of the neurons in the *Drosophila* brain (approximately 1,000 neurons, out of 65,000) involved in parallel groups?
2. Can the neurons in parallel groups be identified in existing work, and related to known cognitive functions?
3. What is the benefit of having a parallel group? Could the same functions not be done by a single larger neuron?
4. Do the different neurons in a group fire synchronously, or do they all have different patterns of firing?
5. Given the diversity of the groups, do the groups have some commonality of function? Or are the groups totally individual?
6. What is the specific function of each group? Can it be related to fruitfly behaviour or other external factors?
7. Why are there so many parallel groups with lateral asymmetry?
8. Why are parallel groups strongly linked to a few of the largest neurons in the *Drosophila* brain, which also have lateral asymmetry?
9. Is there some pattern recognition technique (possibly using AI) which could lead to better identification of parallel groups?

10. Are there parallel groups in the mammalian brain?

For many of these questions, progress depends on experimental work. Is it possible to identify some parallel neurons or their large input-output neurons, and mark them to record single-cell activity with fine time resolution? This would answer some questions, particularly the question of whether parallel neurons fire synchronously, or independently. It would bear on the question of how parallel neurons in a group differ from one another.

It is known that synapses which appear anatomically the same can differ in their synapse strength, for instance by Hebbian learning. Hebbian synapse change is typically thought of as a slow incremental process, requiring large numbers of neural spikes to produce large changes. If the differences between parallel neurons were solely the result of Hebbian changes in synapse strength, we would expect those differences to change only slowly over time. Faster changes in differential parallel firing would point to some other mechanism for differentiating parallel neurons in a group – for instance, multi-state neurons, as suggested in [Worden 2025].

## Appendix: Format of the Supplementary Material

Detailed data about the parallel groups is provided in a comma-separated file groups.csv , which can be viewed as a spreadsheet, for instance using Microsoft Excel.

The first line of the file gives the threshold match value (0.6) used to define the groups, followed by a zero which signifies that groups are defined in terms of the product of input and output matches. There is a single blank line separating each parallel group in the file. The information for each group consists of the following lines:

- A group header line, giving:
    - The number of neurons in the group
    - A name for the group, such as 'lo1', defining its main neuropil
    - The average match of any neuron in the group with all other neurons in the group, in parts per thousand
    - The average reduced match of any neuron in the group with all other neurons in the group, in parts per thousand
    - Following the string 'in:' a list of the short names of neuropils (such as MB_CA_R) where neurons in the group have more than 2% of the input synapses, followed by the percentage of synapses in the neuropil
    - Following the string 'out:' a list of the short names of neuropils where neurons in the group have more than 2% of the output synapses, followed by the percentage of synapses in the neuropil
- A line starting 'descending' giving, for each neuron in the parallel group, its average match with all other neurons in the group, in descending order.
- A line for each neuron in the parallel group, giving:
    - Its full identifier (18 characters)
    - An integer 0, 1, or 2 if it is excitatory, mixed, or inhibitory
    - The rank of the neuron – its position in an ordering of all neurons in the brain, in descending order of their total number of synapses
- A line such as 'Input neuron ids: 464 filtered to those with more than 5 synapses to the group', describing the lines of input neuron ids to follow
- Lines for the input neurons, in descending order of connection to the group. Each line gives the neuron full identifier, followed by detailed information about the neuron and its links to neurons in the group, which can be described on request, followed by the number of synapses from the input neuron to each neuron in the group.
- A line such as 'Output neuron ids: 444 filtered to those with more than 5 synapses to the group', describing the lines of input neuron ids to follow
- Lines for the out neurons, in descending order of connection to the group. Each line gives the neuron full identifier, followed by detailed information about the neuron and its links to neurons in the group, which can be described on request, followed by the number of synapses to the output neuron from each neuron in the group An empty line which terminates the information for the group

## Acknowledgements

I thank Dan Friedman for very valuable comments on an earlier draft.

# References

Abubaker M B, Hsu F, Feng K-L, Chu L-F, de Belle J, Chiang A (2024) Asymmetric neurons are necessary for olfactory learning in the Drosophila brain, Current Biology 34, 946–957

Lee A, Dubuc A et al (2025) Data-driven fine-grained region discovery in the mouse brain with transformers, Nature Communications 16:8536

Lin, A., Yang, R., Dorkenwald, S. *et al.* (2024) Network statistics of the whole-brain connectome of *Drosophila*. *Nature* **634**, 153–165

Pascual, A., Huang, K. L., Neveu, J., & Preat, T. (2004). Neuroanatomy: Brain asymmetry and long-term memory. Nature, 427(6975), 605–606

Wolff T and Rubin G (2018) Neuroarchitecture of the *Drosophila* central complex: A catalog of nodulus and asymmetrical body neurons and a revision of the protocerebral bridge catalog, J Comp Neurol;526:2585–2611.

Worden R (2025) Multi-state neurons, paper presented at the 5th International Conference of AI and Machine Learning, London, and at https://arxiv.org/abs/2512.08815

Zhang T, Zhang X, Sun D and Kim W (2024) Exploring the Asymmetric Body's Influence on Interval Timing Behaviors of Drosophila melanogaster, Behav Genet 54(5):416-425